%% file: hep_ex_dp.tex
\documentstyle[12pt,epsf]{article}
\begin{document}
\Large
\begin{center}
Experimental evidence for a light and broad scalar resonance in 
$D^+\to \pi^-\pi^+\pi^+$ decay .\end{center}
\normalsize
\input{names.tex}
\begin{center}{August, 2000}\end{center}

\begin{abstract}

From a sample of $ 1172 \pm 61 $ $ D^+ \to \pi^- \pi^+ \pi^+ $ decay,
we find
$ \Gamma (D^+ \to \pi^- \pi^+ \pi^+ ) / \Gamma ( D^+ \to K^- \pi^+ \pi^+ ) =
 0.0311 \pm 0.0018 ^{+0.0016}_{-0.0026} $.
Using a coherent amplitude analysis to fit the Dalitz plot of
these decays, we find strong evidence that a scalar resonance
of mass $478^{+24}_{-23} \pm 17 $ MeV/$c^2$ and width
$324^{+42}_{-40} \pm 21 $ MeV/$c^2$  accounts for approximately half of
all decays.
\end{abstract}

The three-body decays of charm mesons often proceed as 
quasi-two-body decays with resonant intermediate states.
In our companion paper\cite{e791ds}, 
we find that $ D_s^+ \to f_0(980) \pi^+ $
accounts for approximately half of the three-pion decay rate and
$ D_s^+ \to f_0(1370) \pi^+ $
accounts for more than half of what remains,
clearly establishing the dominance 
of isoscalar resonances in producing the three-pion final state.
In this paper we present a study of the singly Cabibbo-suppressed
decay $ D^+ \to \pi^- \pi^+ \pi^+ $.
We determine the ratio of decay rates
$ \Gamma (D^+ \to \pi^- \pi^+ \pi^+ ) / \Gamma ( D^+ \to K^- \pi^+ \pi^+ ) $
and study the $ D^+ \to \pi^- \pi^+ \pi^+ $ Dalitz plot to determine the
structure of its density distribution.
We find that allowing an amplitude for an additional scalar state,
with mass and width unconstrained, improves our fit substantially.
The mass and width of the resonance found by this fit are
$ 478^{+24}_{-23} \pm 17  $ MeV/$c^2$ and 
$ 324^{+42}_{-40}  \pm 21$ MeV/$c^2$.
Referring to this $ \pi^+ \pi^- $ resonance as the $ \sigma (500) $,
we find that $ D^+ \to \sigma(500) \pi^+ $ accounts for about
half of the total decay rate.
                    
Other experiments have presented controversial evidence for low-mass
$ \pi \pi $ resonances in partial wave analyses \cite{wa102,cleo,gams},
with ambiguous results for the characteristics of such 
particles\cite {pdg98,torn1}.  Theoretically, 
light scalar and isoscalar resonances are predicted in models
for spontaneous breaking of chiral symmetry such as 
the Nambu-Jona-Lasinio linear $ \sigma $ model \cite{nambu} and its QCD
extension \cite{scadron}. Also, these particles have important consequences 
for the quark model \cite{torn1}, for quark-gluon models \cite{nari}, for understanding low energy $\pi\pi$
interactions \cite{erkelene,penn},  and  for understanding the
$\Delta I = 1/2$ rule\cite{morozumi}.

This study is based on the sample of $ 2 \times 10^{10} $ events recorded 
 in Fermilab experiment E791, in which 500 GeV/$c$~ $ \pi^- $-nucleon 
interactions were observed using  an open geometry spectrometer. The
final analysis makes no direct
use of particle identification; it is solely based on tracking and 
vertex  reconstruction capabilities.  To reduce background, we required a 
3-prong decay  (secondary) vertex to be well-separated from the production 
 (primary) vertex, and located well outside of the target foils and other 
solid material. The  momentum vector of the $D$ candidate had to point back 
to the primary vertex. A more detailed description of the final sample  
selection criteria and some additional details are
  provided in the  companion paper \cite{e791ds} where 
the resulting $\pi^-\pi^+\pi^+$ invariant mass distribution is shown as 
Fig. 1. 

We  have, in addition to the combinatorial
background, three kinds of charm backgrounds: the reflection of the 
$D^+ \to K^-\pi^+\pi^+$ decay, located below 1.85 GeV/c$^2$ in this  spectrum;
 the decay $D^0 \to K^-\pi^+$ 
plus one extra track (mostly from the primary vertex); and the decay chain 
$D_s^+ \to \eta' \pi^+$, $\eta ' \to \rho^0(770) \gamma$, 
$\rho^0(770) \to \pi^+\pi^-$.  The last two reflections
 populate the whole $\pi^-\pi^+\pi^+$ analyzed spectrum. We use Monte Carlo (MC) 
 simulations and data to determine both the shape and the 
size of each type of charm background. The combinatorial background 
  in the $\pi^-\pi^+\pi^+$ invariant mass spectrum is represented
by an exponential function.

We fit the  $ \pi^-\pi^+\pi^+ $ invariant mass distribution 
shown in Fig.1 of reference  \cite{e791ds}
as the sum of $D^+ $ and $ D_s^+ $ 
signals plus background.  
Each signal is described as the sum of two Gaussians with a 
common centroid but different widths, all these parameters
determined by the fit. 
The background is represented by a function with 
four terms described above and in \cite{e791ds}. 
The fit yields 1172 $\pm$  61 
$D^+$ events and 848 $\pm$  44 $D_s^+$ events.

We measure the  branching ratio for $D^+ \to \pi^-\pi^+\pi^+$ 
relative to that of $D^+ \to K^-\pi^+\pi^+$. The $ K^- \pi^+ \pi^+ $ 
signal, selected with the same criteria   used for the 
$ \pi^- \pi^+ \pi^+ $ signal, is 34790 $\pm $  232 events.
The absolute efficiency, $ \varepsilon$, for each decay mode is 
approximately $ 3\% $. From Monte Carlo studies we
determine that the ratio $ \varepsilon(D^+ \to \pi^-\pi^+\pi^+)/
\varepsilon(D^+ \to K^-\pi^+\pi^+)=1.08\pm 0.02$. Note that we use 
the decay matrix element found in this  analysis and an
appropriate one for  $D^+ \to K^-\pi^+\pi^+$ to determine the 
efficiencies. We also weight the MC production
model to match our data. We find the  relative branching ratio to be
\begin{equation}
{ \Gamma(D^+ \to \pi^- \pi^+ \pi^+)  \over{\Gamma(D^+ 
\to K^-\pi^+\pi^+)} } =  0.0311 \pm 0.0018 ^{+0.0016}_{-0.0026}\,. 
\end{equation}
The first error is statistical and the second is systematic. 
Uncertainties in the $ \pi^- \pi^+ \pi^+ $ background shape and 
the levels of some contributions dominate the systematic error.
This result can be compared with the measurements reported by  E691 
\cite{e691}, $0.035 \pm 0.007 \pm 0.003$, by WA82  
\cite{wa82}, $0.032 \pm 0.011\pm 0.003$,  and by E687 \cite{e687},  
$0.043 \pm 0.003 \pm 0.003$.

\begin{figure}[hbt]
\centerline{\epsfysize=3.00in \epsffile{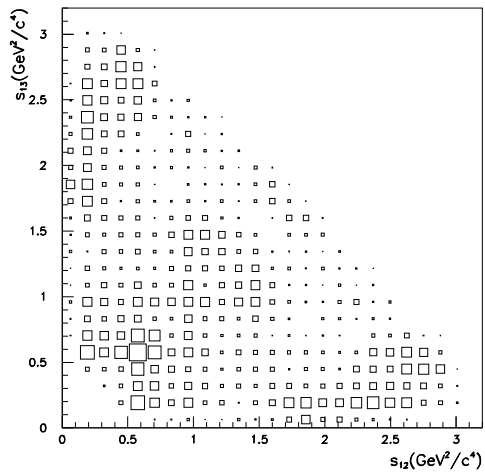}}
\caption  { $ D^+\to \pi^-\pi^+\pi^+ $ Dalitz plot distribution. Since we have 
two identical particles this distribution was symmetrized.}
\label{fig1}
\end{figure}

To study the resonant structure of the decay  
$D^+ \to \pi^- \pi^+ \pi^+$, we consider the 1686  candidates with 
invariant mass between 1.85 and 1.89 GeV. The integrated 
signal-to-background ratio in this range is about 2:1. Fig. \ref{fig1} 
shows the Dalitz plot for these events. The horizontal and vertical 
axes are the squares of the $ \pi^+ \pi^- $ invariant masses, and 
the plot has been symmetrized with respect to the two $ \pi^+ $'s.

To study the resonant structure  in Fig. \ref{fig1},
we use MINUIT\cite{minuit} to extract the parameters. We do this by 
maximizing the log (Likelihood) ${\cal L} $ for several models of signal and 
background. For each model we compute $ {\cal L} $ in terms
of signal and background probability distribution functions (PDF's)  of
the $\pi^-\pi^+\pi^+$ invariant mass, $M$, and the
Lorentz invariants $s_{12} \equiv m_{12}^2$ and $s_{13} \equiv 
m_{13}^2$ (in our convention the odd-charged pion is labeled 
particle $1$). Writing ${\cal P}_S$ and ${\cal P}_B$  for 
the signal and background PDF's, 
${\cal L} =  \prod_{j=1}^{1686} [{\cal P}_S +
{\cal P}_B]_j$. 
We take ${\cal P}_S = 
\frac {1}{N_S} g(M) \varepsilon(s_{12},s_{13})
\mid {\cal A} \mid^2$, with

\begin{equation}
{\cal A} = a_0 e^{i\delta_0}{\cal A}_0 +
\sum_{n=1}^N a_n e^{i\delta_j}{\cal A}_n(s_{12},s_{13}) \, .
\end{equation}
\noindent
In this equation $N_S$ is the normalization constant,
$\varepsilon(s_{12},s_{13})$ is the net efficiency, $g(M)$ is
 a Gaussian function describing the 
signal $\pi^- \pi^+ \pi^+$  invariant mass spectrum.
The fit parameters are the  coefficient magnitudes, 
$a_n$, and the phases, $\delta_n$.

The non-resonant amplitude, $ {\cal A}_0  $, is represented 
by a constant. Each resonant amplitude, ${\cal A}_n( n \ge 1) $, 
is written as a product of four terms.
\begin{equation}
{\cal A}_n = \ ^{J}F_n \times {^J}{\cal M}_n \times BW_n
\end{equation}

The first term is  form factor for the   $n^{th}$ resonance We assume the form factor
for the $D$ decay to be one.
$^{J}{\cal M}_n$ is a term which accounts for angular-momentum 
conservation and depends on the spin J of the resonance.
The final term is a relativistic Breit-Wigner function given by:

\begin{equation}
 BW_n = {1 \over {m^2 - m^2_0 + im_0\Gamma_n(m)}} 
\end{equation}

with

\begin{equation}
\Gamma(m) = \Gamma_0 \frac{m_0}{m}\left(\frac{p^*}{p^*_0}\right)^{2J+1}
\frac{^{J}F_n^2(p^*)}{^{J}F_n^2(p^*_0)}\ .
\end{equation}
In Eqs. 3 and 4, $m$ is the invariant mass of the two pions forming a
spin-J resonance. The functions $^{J}F$ are the Blatt-Weisskopf damping 
factors \cite{blatt}:  $^{0}F = 1$ for spin 0 particles, 
$^{1}F = 1/{\sqrt{1+(rp^*)^2}}$ for spin 1 and 
$ ^{2}F = 1/{\sqrt{9+3(rp^*)^2+(rp^*)^4}}$ for spin 2. The parameter 
$r$ is the radius of the resonance ($\sim 3 fm$) \cite{argus} and 
$p^*=p^*(m)$ the momentum of  decay particles at mass $m$, measured in 
the resonance rest frame, $p^*_0=p^*(m_0)$, where $m_0$ is the 
resonance mass. The spin part of the amplitude 
$^{J}{\cal M}_n$ is defined equal to 1 for a spin-0 resonance,  
${ -2\mid {\bf p_3} \mid \mid {\bf p_2} \mid cos\theta}$ for
spin-1 and ${ \frac{4}{3}(\mid {\bf p_3} \mid \mid {\bf p_2} \mid)^2 
(3cos^2\theta-1)}$, where ${\bf p_3}$ is the 3-momentum of the 
unlike-charge pion and ${\bf p_2}$ is the 3-momentum of the other
like-charge pion, both measured in the resonance rest frame;
and $\theta$ is the angle between pions 2 and 3.
Finally, each resonant amplitude is Bose symmetrized:  
${\cal A}_n = {\cal A}_n[({\bf 12}){\bf3}] + 
{\cal A}_n[({\bf 13}){\bf 2}]$.

The background distribution is given by ${\cal P}_B  = b(M) 
\sum_{i=1}^3 \frac {b_i}{N_{B_i}} {\cal B}_i(s_{12},s_{13})$; 
$b(M)$ is the function describing the background
distribution in the $\pi^- \pi^+ \pi^+$ spectrum,
$b_i$ are the relative amount of each background type and
${N_{B_i}}$ are the corresponding normalization constants.

The three components of the background distribution are the combinatorial 
background, assumed to be uniform before any acceptance effects, and 
the $D^0 \to K^-\pi^+$ and  $D_s^+ \to \eta' \pi^+$ reflections. The
relative background fractions are $80 \pm 6\%$, $4\pm 1\%$, and $16 \pm 6\%$, 
respectively. The shape, location, and size of the charm background 
were obtained using MC simulations and previously determined $D^0$ 
and $D_s$ production rates relative to $D^+$ in our data sample.
All parameters used for the background description 
are fixed during the fit. 

The dominance of the above three background contributions was checked
in several tests.  The analysis was repeated with more stringent 
selection criteria, with various levels of \v{C}erenkov-counter 
requirements on the pions, and with varied background levels in the
fit.  All test results were consistent with the quoted final results.
In addition, Monte Carlo simulations were used to study
specific charm decay channels and a generic sample of charm decays. 
The latter was examined to look for Dalitz-plot structure in the 
events which passed our final selection criteria.  No structures, 
other than those noted above, were significant.

In a first model,
which we will refer to as Fit 1, the signal PDF includes 
a non-resonant amplitude and amplitudes for
$ D^+ $ decaying to a $ \pi^+ $ and  any of five
 established $ \pi^+ \pi^- $ resonances\cite{pdg98}:
$\rho^0(770)$,  $f_0(980)$, $f_2(1270)$, $f_0(1370)$,
 and  $\rho^0(1450)$. In the case of the $f_0(980)$ and $f_0(1370)$, we 
 used the parameters of Ref. \cite{e791ds} and  not those of Ref. \cite{pdg98}. The 
 fit extracts the magnitudes and phases 
of each of the amplitudes along with the error matrix for 
these parameters. 
We calculate the decay fraction for each amplitude as
its intensity, integrated over the Dalitz plot, divided
by the integrated intensity of the signal's coherently summed amplitudes.
The Fit 1 results are listed in  the first column of
Table \ref{fitResults}. In this model, the non-resonant, 
the $\rho^0(1450) \pi^+ $ and the $\rho^0(770) \pi^+ $ amplitudes  
 dominate.
The qualitative features of this fit are similar to
those reported by E691\cite{e691} and E687\cite{e687}.
\begin{center}
   \begin{table}
\centering
 \begin{tabular}{c c c }
\\ \hline     
 & Fit 1&Fit 2  \\  
 mode & Fraction(\%)  &  Fraction(\%)
 \\
  & Magnitude   &  Magnitude
 \\
  &  Phase  &  Phase 
 \\  \hline 
 \vspace*{-10pt} & & \\
  $\sigma\pi^+$  				& --& 46.3 $\pm$ 9.0 $\pm$ 2.1 \\ 
                 				& --&1.17 $\pm$ 0.13 $\pm$0.06    \\
                 				& --&(205.7 $\pm$ 8.0 $\pm$ 5.2)$^{\circ}$  \vspace{.1cm} \\
 $\rho^0(770)\pi^+$ &20.8 $\pm$ 2.4 		&33.6 $\pm$ 3.2 $\pm$ 2.2  \\
                    & 1(fixed)       		& 1(fixed)                 \\ 
                    &0(fixed)        		&0(fixed)  \vspace{.1cm}                \\
		     
   NR              &38.6 $\pm$ 9.7  		&7.8 $\pm$ 6.0 $\pm$ 2.7 \\ 
                   &1.36 $\pm$ 0.20  		&0.48 $\pm$ 0.18 $\pm$ 0.09  \\ 
                   &(150.1 $\pm$ 11.5)$^{\circ}$   &(57.3 $\pm$ 19.5 $\pm$ 5.7)$^{\circ}$ 
\vspace{.1cm}\\
		     		    
  $f_0(980)\pi^+$   &7.4 $\pm$ 1.4  		&6.2 $\pm$ 1.3 $\pm$ 0.4 \\
                    &0.60 $\pm$ 0.07 		& 0.43 $\pm$ 0.05 $\pm$ 0.02  \\ 
                    &(151.8 $\pm$ 16.0)$^{\circ}$  &(165.0 $\pm$ 10.9 $\pm$ 3.4)$^{\circ}$\vspace{.1cm} 
\\
		     
   $f_2(1270)\pi^+$  &6.3 $\pm$ 1.9		&19.4 $\pm$ 2.5 $\pm$ 0.4 \\
                     &0.55 $\pm$ 0.08 		&0.76 $\pm$ 0.06 $\pm$ 0.03  \\ 
                     &(102.6 $\pm$ 16.0)$^{\circ}$ &(57.3 $\pm$ 7.5 $\pm$ 2.9)$^{\circ}$ \vspace{.1cm}\\

   $f_0(1370)\pi^+$ &10.7 $\pm$ 3.1 		&2.3 $\pm$ 1.5 $\pm$ 0.8 \\
                    &0.72 $\pm$ 0.12  		&0.26 $\pm$ 0.09 $\pm$ 0.03 \\ 
                    &(143.2 $\pm$ 9.7)$^{\circ}$  &(105.4 $\pm$ 17.8 $\pm$ 0.6)$^{\circ}$\vspace{.1cm} 
\\

  $\rho^0(1450)\pi^+$ &22.6 $\pm$ 3.7 		&0.7 $\pm$ 0.7 $\pm$ 0.3\\
                      &1.04 $\pm$ 0.12  	&0.14 $\pm$ 0.07 $\pm$ 0.02\\
                      &(45.8 $\pm$ 14.9)$^{\circ}$ &(319.1 $\pm$ 39.0 $\pm$ 10.9)$^{\circ}$\vspace{.1cm} 
\\
  \end{tabular}
 \protect\caption {Final result with the first errors statistical and the  
second, in  Fit 2, the systematic.}
 \label{fitResults}
 \end{table}
\end{center}

\begin{figure}[hbt]
\centerline{\epsfysize=3.8in \epsffile{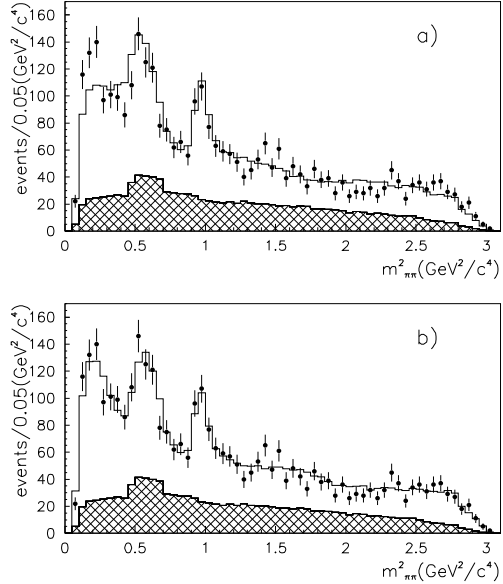}}
\caption{ $s_{12}$ and $s_{13}$ projections for data (error bars) and 
fast-MC (solid). The shaded area is 
the background distribution, (a) solution with the Fit 1, (b)  
solution with Fit 2.} 
\label{fig2}
\end{figure}

To assess the quality of the fit, we developed a fast-MC 
program which produces binned Dalitz-plot  
densities accounting for signal and background PDF's, including
detector efficiency and resolution. Comparing the binned Dalitz-plot-density 
distribution generated by MC events using the magnitudes and 
 phases of the amplitudes given in Fit 1 with that for the data,
we produce the $ \chi^2 $ distribution for the difference 
in densities and observe a concentration of a large  $ \chi^2 $
in the low $ \pi^+ \pi^- $ mass ($m_{\pi^+ \pi^-}$) region. 
The $ \chi^2 $ summed over all 
bins is 254 for 162 degrees of freedom ($ \nu $), which corresponds to
a confidence level less than $10^{-5}$, assuming Gaussian errors. Since the  two $m_{\pi^+ \pi^-}^2$ 
projections are nearly independent, we display the sum of 
$s_{12}$ and $s_{13}$  in  Fig. \ref{fig2}a for the data and 
for the fast-MC. 

The small value of the confidence level casts doubt on the
validity of the model used.
While the projection of the  MC onto the $ \pi^+ \pi^- $
mass${}^2$ axis describes the data in the $ \rho^0 (770) $ and
$ f_0(980) $ regions well, there is a discrepancy at lower mass,
suggesting the possibility of another  amplitude.

To investigate the possibility that another 
$ \pi^+ \pi^- $ resonance contributes
an amplitude to the $ D^+ \to \pi^- \pi^+ \pi^+ $ decay,
we add a sixth resonant amplitude to the signal PDF.
We allow its mass and width to float and assume
a scalar angular distribution.
This fit (Fit 2) converges and finds values 
of $ 478^{+24}_{-23} $  MeV$/c^2$ for the mass
and  $ 324^{+42}_{-40} $ MeV$/c^2$ for the width.
We will refer
to this possible state as the $ \sigma(500) $.
The corresponding results, including the systematic errors \cite{e791ds},
 are collected in the second column 
of Table \ref{fitResults}.

In Fit 2, the $ \sigma (500) $ amplitude produces the largest
decay fraction, 46\%, with a relatively small statistical error,
9\%. The non-resonant fraction, which at $ (39 \pm 10) \% $ was
the largest in the original fit, is now only $ (8 \pm 6 ) \% $.
When we project this model onto the Dalitz plot, the
$ \chi^2 / \nu $ becomes 138/162.
The projection of this model onto the $ \pi^+ \pi^- $ invariant mass
squared distribution, shown in Fig. \ref{fig2}b,
describes the data well, including the accumulation of events near
0.2 GeV$^2/c^4$.
We tested the accuracy of the fit's error estimates by producing hundreds of
fast-MC samples using the parameters of Fit 2 and then fitting
the samples.
The central values of all the parameters are reproduced
accurately, and the width of the $ \sigma(500) $ fraction
distribution is 0.12, slightly larger than MINUIT's estimate, 0.09.

When comparing the two models we used the fast-MC to simulate an ensemble of
samples for each model. For each sample we calculated 
$ \Delta w = -2 ( {\rm ln}\, {\cal L} _i -
{\rm ln} \, {\cal L} _{\sigma} ) $, where $ {\cal L} _{\sigma}$ and ${\cal L}
_i$ are the likelihood functions evaluated with the parameters from the fit with
and without the $\sigma\pi^+$ amplitude, respectively.
In the data, $ \Delta w = 118 $; 
in the fast MC with a $ \sigma $, $ \langle \Delta w \rangle = 108 $;
in the fast MC with no $ \sigma  $,
$ \langle \Delta w \rangle = -106 $.
In both MC experiments, the rms deviation for $ \Delta w $
is about 20,
indicating a strong preference for the additional amplitude,
as is indicated also by the difference in $ \chi^2 / \nu $
for the two  models.

  We consider the systematic errors associated with the values of the
  fixed parameters in the fit.  The most important ones come from
  uncertainties in the background model (the background shape,
  composition, and level), in particular the $D_s^+ \to \eta'\pi^+$ reflection
   which
  populates the same region as the  $D^+ \to \rho^0\pi^+$ component. 
  We also account for the uncertainties in the parameters 
  describing the acceptance function.

To better understand our data,
we also fit it with vector, tensor, and toy models
for the sixth (sigma) amplitude, allowing the masses, widths, and 
relative amplitudes to
float freely.
The vector and tensor models test the angular distribution of the
signal.
The toy model tests the phase variation expected of a Breit-Wigner amplitude
by substituting a constant relative phase.
The vector resonance model converges to poorly defined values of the
mass and width: $ 805 \pm 194 $ MeV/$c^2$ and $ 1438 \pm 903 $ MeV/$c^2 $;
the tensor model to
more poorly defined values:
$ 2350 \pm 683 $ MeV/$c^2 $ and $ -690 \pm 1033 $;
and the toy model to  $ 434 \pm 11 $ MeV/$c^2$ and
$ 267 \pm 37 $ MeV/$c^2 $. As a test of the models, we again project
 the vector, tensor, and toy models onto the Dalitz plot and obtain
  $ \chi^2 / \nu $ = 188/162, 148/162, and $ 152/162 $, respectively.
For these models, we also find
$ \Delta w $ = 66, 13, and 15 where MC
experiments predict $ \langle \Delta w \rangle \approx 64 $, 56,
and 38 when the data is generated with the scalar parameters and
the negatives of those values when the MC data is generated
according to the vector, tensor, and toy model parameters.
The rms widths of the MC distributions are 15, 15, and 11 units
respectively.
These statistical tests strongly exclude the vector model.
They clearly prefer the scalar model to the tensor and toy models.
Note that the central value for the tensor mass is well above threshold
for $ D^+ $ decay and the negative width is an indication that no physically
meaningful tensor resonance fits the data.
In the toy model, the extra amplitude interferes strongly with a
large non-resonant amplitude, leading to an unphysically large sum
of resonant fractions.

In summary, from 1172 $\pm$  61
$D^+\to \pi^- \pi^+ \pi^+$ we have measured 
$\Gamma(D^+ \to \pi^- \pi^+ \pi^+)/\Gamma(D^+ \to K^- \pi^+ \pi^+)
 = 0.0311 \pm 0.0018 ^{+0.0016}_{-0.0026}$. 
In an amplitude analysis of a sample with S:B $ \approx $ 2:1
we find strong evidence that a scalar resonance with mass
$478^{+24}_{-23} \pm 17 $ MeV/$c^2$ and width
$324^{+42}_{-40} \pm 21 $ MeV/$c^2$
produces a decay fraction $ \approx 50\% $.
Alternative explanations of the data fail to describe
it as well.
The prominence of an amplitude for an isoscalar plus a $\pi^+ $ 
in this decay accords well with our observation
that the amplitude for an isoscalar plus a pion  produce a large 
majority of all $ D_s^+	 \to \pi^- \pi^+ \pi^+ $ decays\cite{e791ds}.

We gratefully acknowledge the assistance of the staffs of Fermilab and of all
the participating institutions.  This research was supported by the Brazilian
Conselho Nacional de Desenvolvimento Cient\'{\i}fico e Tecnol\'{o}gico,
CONACyT (Mexico), the Israeli Academy of Sciences and Humanities, 
the U.S. Department of Energy, the U.S.-Israel
Binational Science Foundation, and the U.S. National Science Foundation.
Fermilab is operated by the Universities Research Association, Inc., under
contract with the U.S. Department of Energy.

\small
\bibliographystyle{unsrt}

\end{document}

%% file: names.tex
\small\begin{center}
    E.~M.~Aitala,$^9$
       S.~Amato,$^1$
    J.~C.~Anjos,$^1$
    J.~A.~Appel,$^5$
       D.~Ashery,$^{14}$
       S.~Banerjee,$^5$
       I.~Bediaga,$^1$
       G.~Blaylock,$^8$
    S.~B.~Bracker,$^{15}$
    P.~R.~Burchat,$^{13}$
    R.~A.~Burnstein,$^6$
       T.~Carter,$^5$
    H.~S.~Carvalho,$^{1}$
    N.~K.~Copty,$^{12}$
    L.~M.~Cremaldi,$^9$
       C.~Darling,$^{18}$
       K.~Denisenko,$^5$
       S.~Devmal,$^3$
       A.~Fernandez,$^{11}$
    G.~F.~Fox,$^{12}$
       P.~Gagnon,$^2$
       C.~Gobel,$^1$
       K.~Gounder,$^9$
    A.~M.~Halling,$^5$
       G.~Herrera,$^4$
       G.~Hurvits,$^{14}$
       C.~James,$^5$
    P.~A.~Kasper,$^6$
       S.~Kwan,$^5$
    D.~C.~Langs,$^{12}$
       J.~Leslie,$^2$
       B.~Lundberg,$^5$
       J.~Magnin,$^1$       
       A.~Massafferri,$^1$
       S.~MayTal-Beck,$^{14}$
       B.~Meadows,$^3$
 J.~R.~T.~de~Mello~Neto,$^1$
       D.~Mihalcea,$^7$
    R.~H.~Milburn,$^{16}$
    J.~M.~de~Miranda,$^1$
       A.~Napier,$^{16}$
       A.~Nguyen,$^7$
    A.~B.~d'Oliveira,$^{3,11}$
       K.~O'Shaughnessy,$^2$
    K.~C.~Peng,$^6$
    L.~P.~Perera,$^3$
    M.~V.~Purohit,$^{12}$
       B.~Quinn,$^9$
       S.~Radeztsky,$^{17}$
       A.~Rafatian,$^9$
    N.~W.~Reay,$^7$
    J.~J.~Reidy,$^9$
    A.~C.~dos Reis,$^1$
    H.~A.~Rubin,$^6$
    D.~A.~Sanders,$^9$
 A.~K.~S.~Santha,$^3$
 A.~F.~S.~Santoro,$^1$
       A.~J.~Schwartz,$^{3}$
       M.~Sheaff,$^{17}$
    R.~A.~Sidwell,$^7$
    A.~J.~Slaughter,$^{18}$
    M.~D.~Sokoloff,$^3$
       J.~Solano,$^1$
    N.~R.~Stanton,$^7$
    R.~J.~Stefanski,$^5$  
       K.~Stenson,$^{17}$ 
    D.~J.~Summers,$^9$
       S.~Takach,$^{18}$
       K.~Thorne,$^5$
    A.~K.~Tripathi,$^{7}$
       S.~Watanabe,$^{17}$
 R.~Weiss-Babai,$^{14}$
       J.~Wiener,$^{10}$
       N.~Witchey,$^7$
       E.~Wolin,$^{18}$
    S.~M.~Yang,$^7$
       D.~Yi,$^9$
       S.~Yoshida,$^7$
       R.~Zaliznyak,$^{13}$ and
       C.~Zhang$^7$ \\
\normalsize (Fermilab E791 Collaboration)\\
\small
{\it 
$^1$ Centro Brasileiro de Pesquisas F{\'\i}sicas, Rio de Janeiro, Brazil,
$^2$ University of California, Santa Cruz, California 95064,
$^3$ University of Cincinnati, Cincinnati, Ohio 45221,
$^4$ CINVESTAV, Mexico City, Mexico,
$^5$ Fermilab, Batavia, Illinois 60510,
$^6$ Illinois Institute of Technology, Chicago, Illinois 60616,
$^7$ Kansas State University, Manhattan, Kansas 66506,
$^8$ University of Massachusetts, Amherst, Massachusetts 01003,
$^9$ University of Mississippi-Oxford, University, Mississippi 38677,
$^{10}$ Princeton University, Princeton, New Jersey 08544,
$^{11}$ Universidad Autonoma de Puebla, Puebla, Mexico,
$^{12}$ University of South Carolina, Columbia, South Carolina 29208,
$^{13}$ Stanford University, Stanford, California 94305,
$^{14}$ Tel Aviv University, Tel Aviv, Israel,
$^{15}$ Box 1290, Enderby, British Columbia, V0E 1V0, Canada,
$^{16}$ Tufts University, Medford, Massachusetts 02155,
$^{17}$ University of Wisconsin, Madison, Wisconsin 53706,
$^{18}$ Yale University, New Haven, Connecticut 06511
}\end{center}
\normalsize